# Role of Magnetism in Superconductivity of BaFe$_2$As$_2$: Study of 5$d$ Au-doped Crystals

Li Li,[1] Huibo B. Cao,[2] Michael A. McGuire,[1] Jungsoo S. Kim,[3] Greg R. Stewart,[3] and Athena S. Sefat[1]

[1] *Materials Science & Technology Division, Oak Ridge National Laboratory, Oak Ridge, TN 37831*
[2] *Quantum Condensed Matter Division, Oak Ridge National Laboratory, Oak Ridge, TN 37831*
[3] *Physics Department, University of Florida, Gainesville, FL 32611*

*Corresponding author emails:* lil2@ornl.gov; sefata@ornl.gov

We investigate properties of BaFe$_2$As$_2$ (122) single crystals upon gold doping, which is the transition metal with the highest atomic weight. The Au substitution into the FeAs-planes of 122 crystal structure (Au-122) is only possible up to a small amount of ~3%. We find that 5$d$ is more effective in reducing magnetism in 122 than its counter 3$d$ Cu, and this relates to superconductivity. We provide evidence of short-range magnetic fluctuations and local lattice inhomogeneities that may prevent strong percolative superconductivity in Ba(Fe$_{1-x}$Au$_x$)$_2$As$_2$.

## I. INTRODUCTION

High-temperature superconductivity (HTS) is among the most mysterious and elusive properties in condensed matter physics, which has now been unveiled in two Cu- and Fe-based families. Many transition-metal based, tetragonal structures with layers have attracted attention following the discovery of iron-based superconductors (FeSC) in LaFeAsO, an antiferromagnetic spin-density-wave (AF SDW) material.[1] The FeSC share some common features with the cuprate family,[2,3] and most importantly it seems that HTS is triggered by chemical doping (or pressurizing) of an AF 'parent' material.[2–5] The parents of FeSC are itinerant weakly-correlated poor metals,[6] with a Fermi surface that is sensitive to small changes in composition,[7–9] and can even tolerate in-plane disorder.[10] In fact, small substitution of Fe with Co can be described by the simple shift of Fermi energy for the one additional electron (assuming +2 ions).[10] Despite the rich chemistry that FeSC offers[11] and the vast experimental and theoretical work exemplified here and through many review manuscripts,[12–17] many things about them (*e.g.*, doping trends, HTS and $T_C$ values) remain a conundrum.

BaFe$_2$As$_2$ ('122') is a parent of FeSC that transitions from the tetragonal (*I*4/*mmm*) non-magnetic state into the orthorhombic (*Fmmm*) SDW striped-AF phase reported below $T_N$= 140 K (polycrystals),[18] $T_N$= 132 K (FeAs-grown crystals),[19] or even 85 K (in Sn-grown crystals)[20]. Such experimental differences in transition temperatures generally depend on sample quality (impurities, flux substitution), off-stoichiometries (*e.g.*, vacancies), and structural details (mixed atomic occupancies, local atomic clustering). For example, we have recently demonstrated that CaFe$_2$As$_2$ crystals can hold complex local structural differences and bond displacements that dictate their property variations.[21-23] Here we produce Au-122 crystals, using our typical self-flux technique.[10,11,19] For 122, in-plane (FeAs *ab*-plane) transition-metal doping with either holes (*e.g.*, 3*d* Cr, Mn; 4*d* Mo)[24–27] or electrons (*e.g.*, 3*d* Co, Ni; 4*d* Rh, Pd)[10,28–30] suppresses AF, but only electron dopants can instigate superconductivity. The reason for the latter are not exactly solved, especially since the dopants can be very low in concentration. In addition, it is found that electron-doping of 122 using 3*d* or 4*d* in the same group (Co and Rh, or Ni and Pd) give overlapping temperature-composition (*T*-x) phase diagrams,[31] *i.e.*, they present the same rate of $T_N$ suppression, the maximum $T_C$, and the range of the superconducting dome. However, this trend breaks for 5*d*; for example, Pt-doping is reported to give $T_C$ in 122 at smaller x ≈ 0.01 and shows much wider x superconducting region (x ≈ 0.01 to 0.11),[32,33] while Ir-122 has $T_{C,max}$ (28 K) for x = 0.15,[34] in marked contrast to Co- or Rh-doping in same group, for which $T_C$ is reduced to less than 10 K for the same x. In this study we chemically substitute Au within FeAs layers of 122, which is another 5*d* element but with the highest atomic weight among transition metals.

Compared to nominal $Fe^{2+}$ ($3d^6$) in 122, Au substitution may signify addition of electrons ($Au^+$: $d^{10}$, $Au^{3+}$: $d^8$) and expansion of the crystal structure due to its extended orbitals, which is noted by transition metal-arsenide bond lengths of ~ 2.40 Å in 122,[35] and ~ 2.74 Å in $LaAuAs_2$, which has similar tetrahedral coordination around the transition metal.[36] In the periodic table, Au sits to the right of 5*d* Ir and Pt, and is just below Cu (3*d*) and Ag (4*d*). Although there are no studies on Ag-122 (presumably because it can be +1, may not form a coordination with As, and is too large to sit in interstitial sites), there was a doping study of Cu into 122.[29] Thermoelectric power and Hall coefficient data give evidence for a similar change of electronic properties for both Co- or Cu-doping of 122 at comparable *e* values (nominal extra dopant electrons) close to that associated with superconductivity,[29,37] even though Co-122 has larger superconducting dome ($T_{C,max}$≈22 K, and Δx =0.1) than Cu-122 ($T_{C,max}$=2 K, and Δx =0.015). Based on this, it is deduced that the establishment of a proper *e* value is not a sufficient condition for superconductivity.[27] Moreover, it is found that although Co-122 can be described by the rigid band picture,[10,38] the total extra electron number estimated from the Fermi surface volumes decreases in going from Co-, to Ni-, to Cu-122, described by increasing impurity potential.[39] Most recently, our nuclear magnetic resonance results for Cu-122 attribute the absence of the large superconducting dome in the phase diagram of Cu-122 to the emergence of a nearly magnetically ordered FeAs plane under the presence of orthorhombic distortion.[40] In fact, the strength of spin fluctuations ($1/T_1T$), where $T_1$ is the $^{75}$As nuclear spin-lattice relaxation rate, remains high for Cu-122, even though greatly reduced upon Co doping.[40] In this work we find that 5*d* Au-doping causes faster decrease in $T_N$ compared with 3*d* Cu-122, which seems to shift the superconducting region to lower x. However, the rate of drop of the structural transition (with x) closely follows Cu-122. We also show evidence of dopant non-uniformity and short-range scale magnetism that may prevent bulk superconductivity.

## II. RESULTS AND DISCUSSION

Single crystals of Au-doped $BaFe_2As_2$ were grown out of self-flux using a high-temperature solution-growth technique.[11] To produce a range of dopant concentrations, small barium chunks, gold pieces, and FeAs powder were combined according to various loading ratios of Ba:Au:FeAs = 1:x:4 (listed in Table 1) in a glove box, and each placed in an alumina crucible. A second catch crucible containing quartz wool was placed on top of this growth crucible and both were sealed inside a silica tube under ~1/3 atm argon gas. Each reaction was heated for ~24 h at 1180 °C, and then cooled at a rate of 1 to 2°C/h, followed by a decanting of the flux between 1090 and 1030 °C. The crystals were flat with dimensions of ~6 × 4 × 0.1 mm$^3$ or smaller. Similar to 122,[19] the crystals of Au-122 formed with the [001] direction perpendicular to the flat faces. Attempts for higher Au contents were unsuccessful and only led to phase separation and other phases. The chemical composition of crystals was measured with a Hitachi S3400 scanning electron microscope operating at 20 kV; energy-dispersive x-ray spectroscopy (EDS) indicated that significantly less Au is chemically-substituted in the 122 structure than put in solution. Three spots (~ 80 μm) were checked and averaged on each crystal; no impurity phases or inclusions were detected. It is assumed that Au sits on the Fe site as there is small deficiency of Fe upon Au-doping. The samples are denoted by these measured EDS x values in Ba(Fe$_{1-x}$Au$_x$)$_2$As$_2$ throughout this paper (Table 1); the error on x is on the order of 5%.

**Table 1**: For $Ba(Fe_{1-x}Au_x)_2As_2$, loading reaction ratio, gold amount found from EDS; room-temperature lattice parameters refined from x-ray diffraction data.

| Au:    | x     | c (Å)       | a (Å)      |
|--------|-------|-------------|------------|
| 0 : 4    | 0     | 13.0151(3)  | 3.9619(2)  |
| 0.05 : 4 | 0.005 | 13.0163(2)  | 3.9626(2)  |
| 0.10 : 4 | 0.009 | 13.0176(3)  | 3.9646(3)  |
| 0.20 : 4 | 0.012 | 13.0186(2)  | 3.9669(2)  |
| 0.30 : 4 | 0.031 | 13.0208(1)  | 3.9705(1)  |

Bulk phase purity of Au-122 crystals was checked by collecting data on an X'Pert PRO MPD X-ray powder diffractometer in the 10-70° 2θ range, on ground crystals weighing ~ 30 mg collectively. Lattice parameters were refined from full-pattern LeBail refinements using the program FULLPROF. The Bragg reflections were indexed using the tetragonal ThCr$_2$Si$_2$ tetragonal structure (*I4/mmm*) without any contributions from impurity phases. Fig. 1a shows the typical diffraction pattern, here for x = 0.031, with good Rietveld refinement ($R_{wp}$ = 7.5%). The refined lattice constants are listed in Table 1; Fig. 1b plots *a*- and *c*- lattice parameters as a function of Au value, and depicts cell volume expansion with larger 5*d* (inset). With small Au doping, *a*- and *c*- lattice parameters increase slightly and monotonically; for 3.1% chemical substitution, the overall unit cell volume expands less than 1 % (~ 0.5%). The arsenic height from Fe layer ($z_{As}$= 0.365 Å), refined from room temperature data and upon Au substitution from x = 0 to 0.031, only changes by less than 1 %.

Magnetization measurements were performed in the Quantum Design magnetic property measurement system upon warming in a magnetic field. Fig. 2a and b present the magnetic susceptibility measured along *ab*- and *c*-crystallographic directions. For BaFe$_2$As$_2$, the susceptibility decreases approximately linearly with decreasing temperature, then drops abruptly below $T_N$=$T_O$≈132 K reproducing the well-established behavior.[19] There is a small anisotropy as $\chi_{ab}$(380 K) = 1.03×10$^{-3}$ cm$^3$/mole and $\chi_c$(380 K)= 0.82×10$^{-3}$ cm$^3$/mole. For all Au-

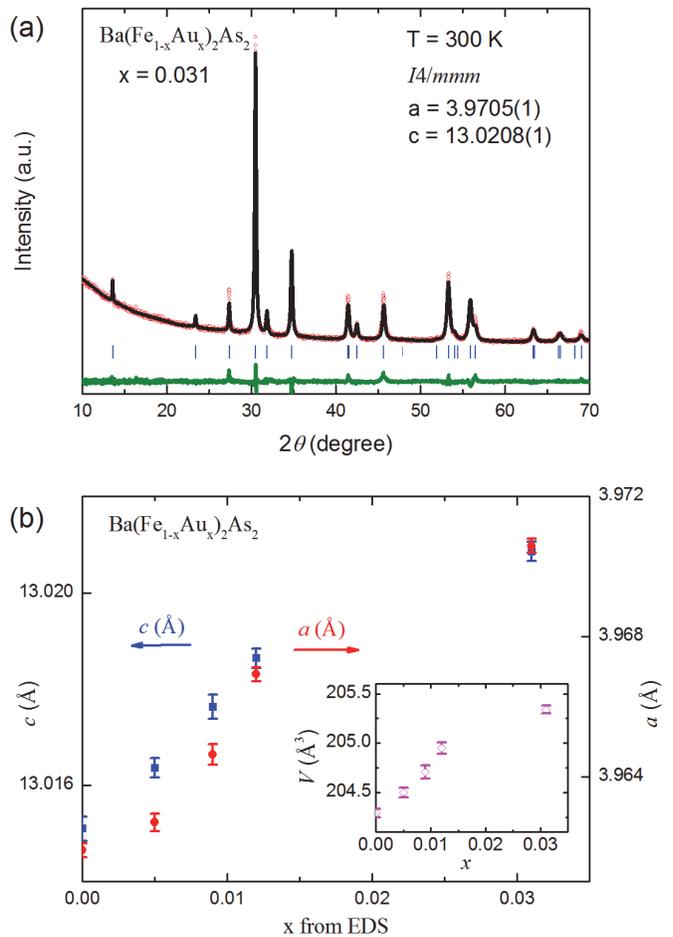

**Fig. 1:** For Au-122: (a) room-temperature powder x-ray diffraction pattern for x=0.031. Red circles represent observed data; black and green solid lines represent the calculated intensity and difference between the observed and calculated intensity; blue vertical bars indicate the Bragg reflection positions; (b) refined lattice parameters for 0 ≤ x ≤ 0.031, inset is cell volume *V* versus x.

122 and above ~ 150 K, the susceptibility data nearly overlap with comparable linear dependence. The magnetic anomaly is reduced in temperature with x. For x = 0.031, $\chi$ increases below the transition temperature indicating additional magnetic contributions. For x = 0.005, 0.009, 0.012, and 0.031, $T_N$ values are inferred as ≈ 121 K, 113 K, 97 K, and 64 K, respectively, using Fisher's $d(\chi T)/dT$.[41] For x = 0.031, enlarging 1 T data shows a small kink in $\chi_{ab}$ around 10 K (Fig. 2c). Despite the larger overall magnetization value, diamagnetic signal is obtained only for this composition at 10 Oe (Fig. 2c inset), with divergence of cooled/warmed data suggesting a superconducting contribution below 2.5 K.

The electrical transport and heat capacity measurements down to 1.8 K were performed in a Quantum Design physical property measurement system. Electrical leads were attached to the crystals using Dupont 4929 silver paste and resistance measured in the *ab* plane in the range of 1.8 to 380 K. The resistivity at 380 K ranged from 0.1 to 1.2 mΩ cm for all x in Au-122. Fig. 3a presents normalized $\rho/\rho_{380K}$; in inset, each x is shifted upward by 0.3 to clarify anomalies. Electrical resistivity for 122 is as expected, and the anomaly is suppressed monotonically with increasing x similar to literature.[19,24-30] For lightly Au-doped composition of x = 0.005 and 0.009, sharp features occur around 122 and 112 K, respectively. The resistivity for x ≥ 0.012 first decreases gently from 380 K, followed by sharp upturns below 102 K for x = 0.012, and 64 K for x = 0.031. Such upturns and continued increase of ρ with decreasing temperature are similar to what occurs in other electron-doped crystals.[10,28–30] The upturn reflects the loss of carriers as a partial SDW gap opens below $T_N$. At temperatures well below $T_N$, the increase in the mobility of the remaining carriers is not enough to overcome the lower carrier concentration and the resistivity continues to increase. Fig. 3b displays ρ(T) from 1.8 to 100 K for the 3.1% doped crystal. ρ(T) slowly increases with cooling, passes through a broad increase at 64 K (defined by $d\rho/dT$), followed by a drop below ~6 K, reaching zero at ~ 2 K. The field dependence of this transition (see inset) is consistent with the diamagnetic $\chi$ signal (Fig. 2c), for evidence of superconductivity. The broad $\Delta T_C$ may signify local chemical inhomogeneity. Hall coefficient ($R_H$) data for 0.012 and 0.013 are presented in Fig. 3c. $R_H$ of 122 is negative in the whole temperature region of 10 to 300 K, a sign of dominant electron contribution, with a sharp decrease below structural/magnetic transition near 132 K. The values of $R_H$ for x > 0 are also negative between 10 and 300 K, with features at 110 K for x = 0.012, and ~70 K for x = 0.031, consistent with Fermi surface gapping scenario for $T_N$. The overall change of Hall data for x = 0.012 and 0.031 are not as rapid as 122, which signify a weaker electronic structure change and reduced magnetism. The widths of transitions for x = 0.031 is more broad and also $R_H$ values fall between x = 0 and 0.012. The local lattice strain and phase coexistence due to non-uniform chemical substitution, for such small doping levels, may cause such effects giving more contributions from electron sheets.

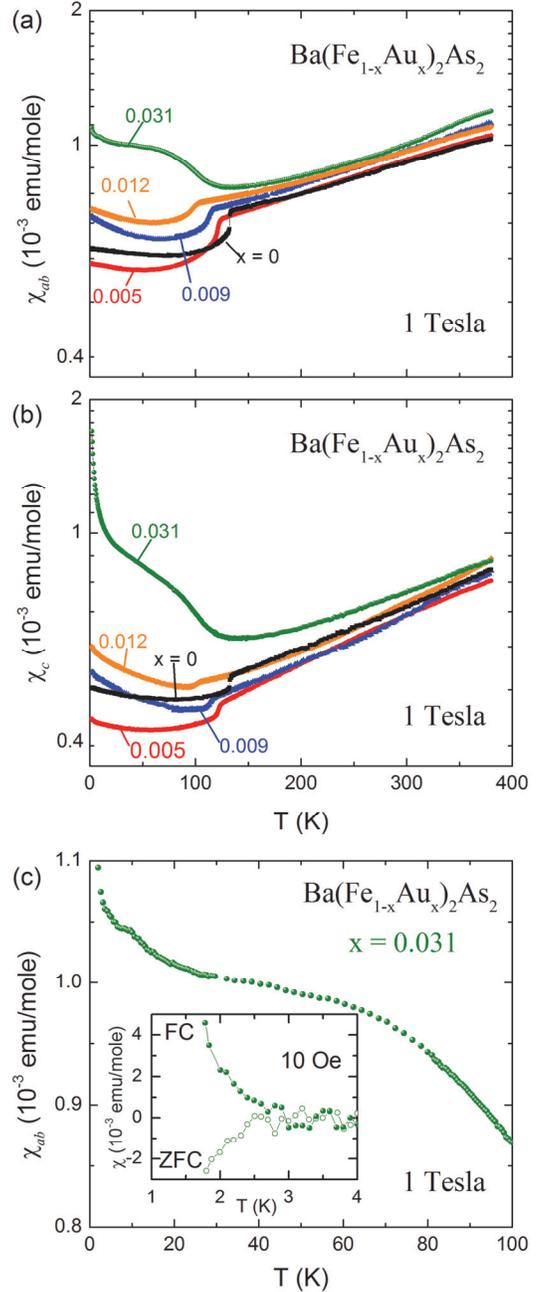

**Fig. 2:** For Au-122, 0 ≤ x ≤ 0.031, (a) temperature dependence of magnetic susceptibility (a) along *ab*-, and (b) *c*-lattice directions. (c) $\chi(T)$ behavior enlarged for x= 0.031 below 100 K at 1 T; the inset is data below 4 K taken at 10 Oe.

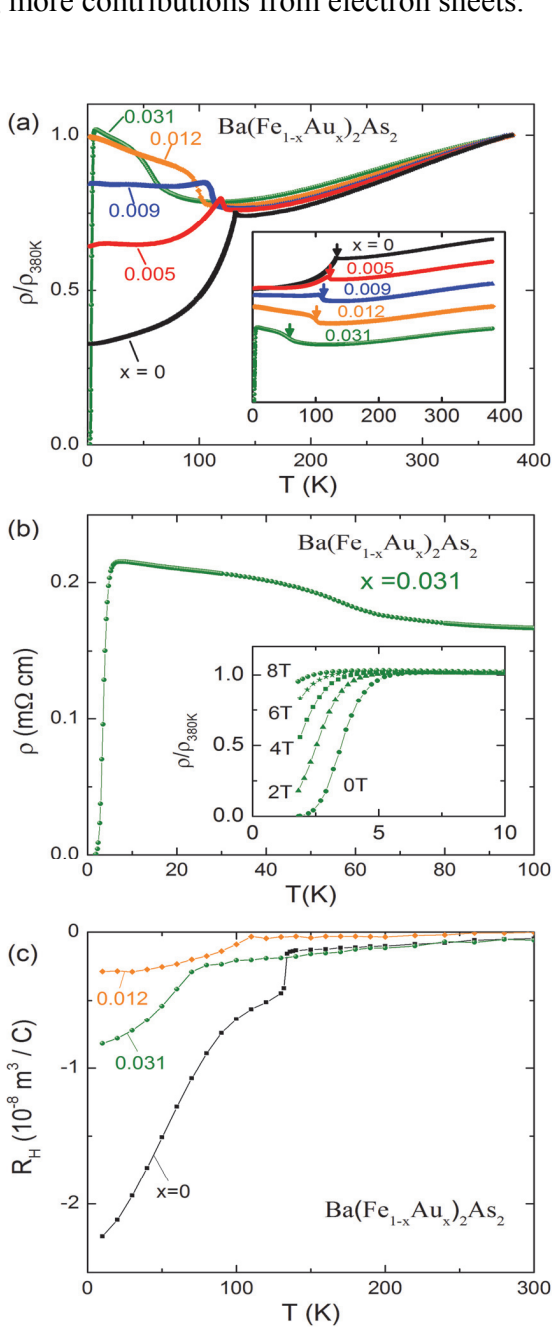

**Fig. 3:** For Au-122, temperature dependent resistivity for (a) $0 \leq x \leq 0.031$ normalized to 380 K (inset has arbitrary ρ). (b) ρ(T) for x= 0.031 below 100 K, with field dependence in inset. (c) Hall coefficient for x = 0, 0.012 and 0.031.

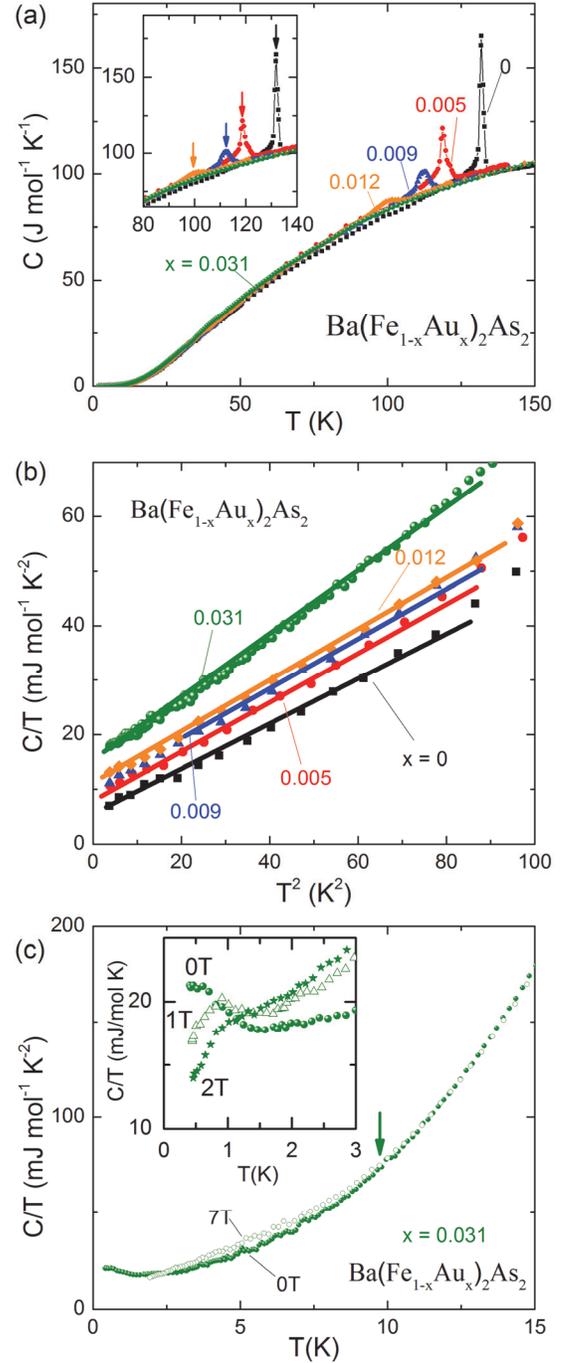

**Fig. 4:** For Au-122, heat capacity for $0 \leq x \leq 0.031$ (a) below 150 K with inset showing the enlarged data around transitions; (b) in form of $C/T$ versus $T^2$ below 10 K; (c) below 15 K at applied fields for x = 0.031 with inset measuring below 3 K.

Heat capacity data are shown in Fig.4. For 122, a sharp transition is observed at 132 K, as expected, for overlapping $T_N$ and $T_O$. With Au doping, the peak decreases monotonically (Fig. 4a): for x = 0.005, 0.009, and 0.012, the tops occur at 118.7 K, 112.2 K, and 100.4 K, respectively. With Au doping, the peaks broaden significantly too (see inset of Fig. 4a) without sharp characteristics, signifying phase inhomogeneity. For x = 0.031, there are no contribution in heat capacity at ~ 64 K, as was seen in $\chi(T)$ and $\rho(T)$, suggesting short-range magnetism or magnetic fluctuations. The Sommerfeld coefficient $\gamma$ for all x is estimated between ~6 to 16 mJ.mol$^{-1}$.K$^{-2}$ (Fig. 4b). This weak change in $\gamma$ with x is similar to that observed for Ba(Fe$_{1-x}$Mo$_x$)$_2$As$_2$,[25] as may be expected for such low-doping levels. For x = 0.031, the zero field and 7 Tesla heat capacity data split near 10 K, which is consistence with the anomaly observed in $\chi_{ab}$ (Fig. 2c). This may be associated with the formation of in-plane local magnetic order that needs to be confirmed by further studies and through other techniques such as neutron scattering. The low temperature heat capacity data (taken in a self-made calorimeter) only shows a Schottky-like feature below 2 K (the inset of Fig. 4c), with no bulk superconductivity transition evident. However, note that in FeSC, the expected size of $\Delta C/T_C$ (from the correlation between $\Delta C/T_C$ and $T_C$)[42] for a $T_C$ of 2-2.5 K would be only 0.5 mJ/mol.K$^2$, or 3% of the measured C/T at this temperature as shown in the inset in Fig. 4c. This is consistent with the weak nature of superconductivity for x = 0.031 Au-122 crystal. Our preliminary room-temperature TEM images show some signs of local crystal lattice strain with ~ 1% Au-doping that may support such broadened transitions.

Single crystal neutron diffraction was performed on the crystal with x = 0.005 (~0.02 g), measured at the four-circle diffractometer HB-3A at the High Flux Isotope Reactor at ORNL. The neutron wavelength of 1.542 Å was used from a Si-220 monochromator.[43] Results are shown in Fig. 5a and b. The order parameter to the SDW order is seen by the intensity of the magnetic reflection (½ ½ 5)$_T$, presented in the tetragonal cell. For tracking the tetragonal-to-orthorhombic transition, the intensity of the (2 2 0)$_T$ nuclear peak was measured with warming; the intensity increase below the $T_O$ is due to a reduced extinction effect caused by the structural transition. To further confirm the structural transition, we also performed low temperature powder x-ray diffraction. The angular range near the tetragonal (1 1 2) reflection [orthorhombic (2 0 2) and (0 2 2)] was carefully examined at different temperatures. Fig. 5c shows that the single peak of (1 1 2) gradually broadens and finally splits into two peaks as the sample is cooled through the symmetry-lowering crystallographic phase transition.

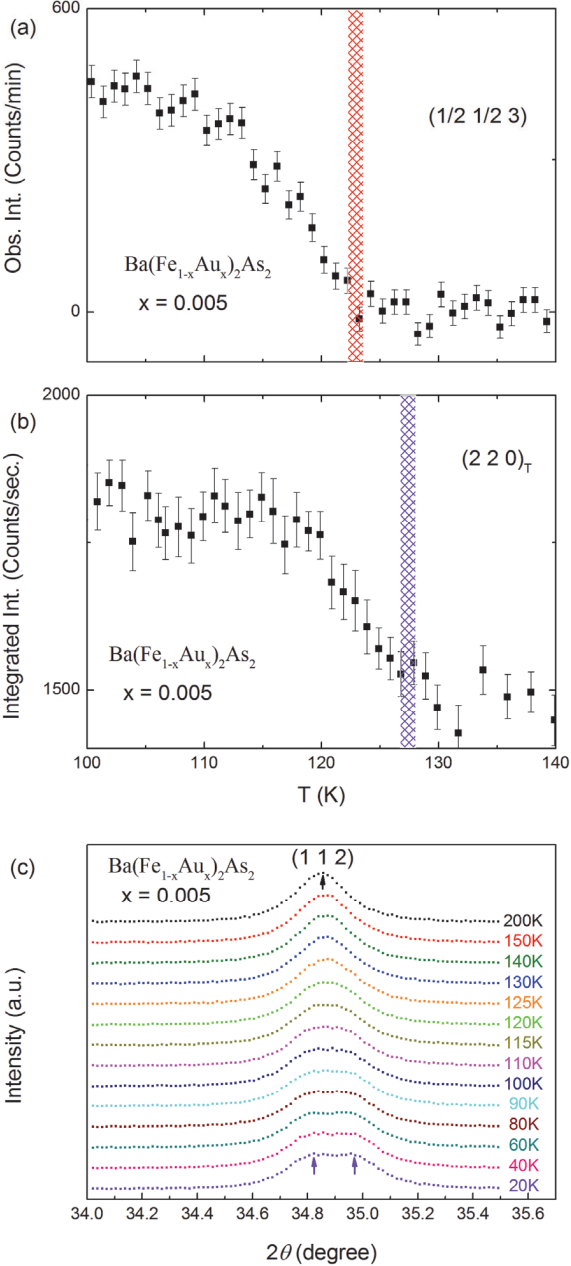

**Fig. 5:** For Au-122 with x = 0.005, (a, b) the neutron (c) and the X-ray diffraction data. The temperature-dependence of magnetic (½ ½ 5)$_T$ and nuclear (2 2 0)$_T$ reflections, gives the onset of the AF transition below $T_N$ =122 K, and orthorhombic transition below $T_O$ = 128 K. (c) Tracking the (1 1 2) reflection with temperature, clearly broadens with weak splitting to the orthorhombic (2 0 2) and (0 2 2) evident at 20 K.

In conclusion, this work investigated the Au-doping effects on $BaFe_2As_2$ single crystals for the first time. The T-x phase diagram can be constructed for the $Ba(Fe_{1-x}Au_x)_2As_2$ system, shown in Fig. 6. The suppression rate of the $T_N$ with x is faster than that reported for $3d$ Cu-122, indicating that Au is more disruptive than Cu.[29] For x = 0.031, weak superconductivity with $T_C$ ~ 2 K and anomalies near ~ 64 K in $\chi$ and $\rho$ may signify short-range magnetic correlations in the nematic region. Also, a weak anomaly occurs near 10 K in $\chi_{ab}$ for x = 0.031, correlated with splitting between 0 and 7 T in $C$ data that may also be related to magnetism. The broadened $C$ transitions may indicate local lattice strain and chemical non-uniformity, which may lead to ordered FeAs planes similar to that seen in Cu-122, ultimately preventing higher temperature superconductivity. This study demonstrates the close relationship between materials' structural details such as dopant types/concentrations and potential clustering/inhomogeneity in causing temperature-dependent phase transformations.

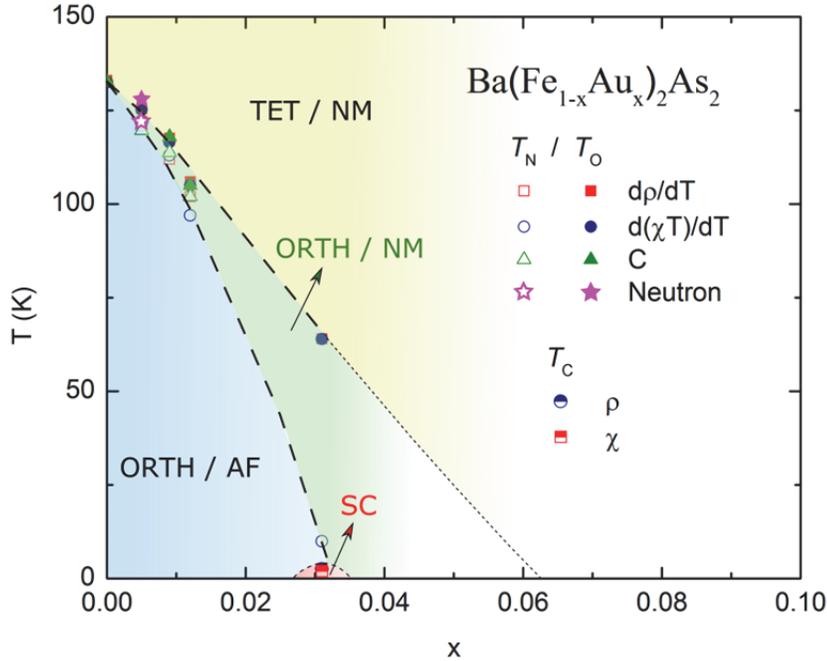

**Fig. 6**: T-x phase diagram for Au-122.

This work was primarily supported by the U. S. Department of Energy (DOE), Office of Science, Basic Energy Sciences, Materials Science and Engineering Division (AS, MM). This study was partially funded (LL) by ORNL's Lab-directed Research & Development (LDRD). The work at ORNL's HFIR (HC) was sponsored by the Scientific User Facilities Division, Office of Basic Energy Sciences, U.S. Department of Energy. The work at University of Florida was funded by the U. S. DOE, Office of Basic Energy Sciences, contract no. DE-FG02-86ER45268. We acknowledge B.C. Chakoumakos and T. Imai for fruitful scientific discussions and their input on this manuscript. We appreciate A.F. May and J. Yan for technical support.